\documentclass[preprint]{jpsj2}

\def\runtitle{Numerical Study of Excited States
 in the Shastry-Sutherland Model}
\def\runauthor{Tomo  \textsc{Munehisa} and Yasuko \textsc{Munehisa}}

\title{%
Numerical Study of Excited States in the Shastry-Sutherland Model
}

\author{%
Tomo  \textsc{Munehisa} and 
Yasuko \textsc{Munehisa}
}

\inst{%
 Faculty of Engineering, Yamanashi Univ., Kofu 400-8511
}

\recdate{\today}

\abst{%

We investigate excited states of the
Shastry-Sutherland model using a kind of variational method.
Starting from various trial states which include one or two triplet dimers, 
we numerically pursue the best evaluation of the energy  for 
each set of quantum numbers.
We present the energy difference as a function of either the coupling 
ratio or the momentum and compare them with
the perturbative calculations. 
Our data suggest that the helical order phase exists
between the singlet dimer phase and the magnetically ordered phase.
In comparison with the experimental data we can estimate
the intra-dimer coupling $J$ and the inter-dimer coupling $J'$ for
 $\rm{ SrCu_2(BO_3)_2 }$: $J'/J =0.65$ and $J = 87$K.
}

\kword{%
Shastry-Sutherland model, excited states, variational method,
dispersion, helical order
}

\begin{document}
\sloppy
\maketitle

\section{Introduction}

One of recent fascinating topics in two-dimensional quantum spin systems is 
the Shastry-Sutherland model (SS model)\cite{SS}, 
the model of orthogonal dimers with the intra-dimer coupling $J \ ( > 0)$ 
and the inter-dimer coupling $J'\ ( > 0)$. 
This model provides a nontrivial system which relates an  
exactly solvable spin-gaped model obtained in the $J'\rightarrow 0$ limit   
to a square lattice model where intra-dimer interaction vanishes ($J = 0$).
It is known that the exact ground state of the SS model is the direct 
product of the singlet dimers when $J'\le J'_c \sim 0.68J$, while  
for the sufficiently large $J'/ J$ the system is in the 
magnetic order phase with the N\'{e}el ground state.

A lot of notable studies have been done on this system since 
Kageyama~$et$~$al.$\cite{exp1} found that $\rm{ SrCu_2(BO_3)_2 }$ realizes
the SS model and its coupling ratio $J'/J$ is very close 
to the critical value of the model. 
Active experimental works have reported additional interesting
features of this matter such as 
magnetization  plateaus \cite{exp1,plateau,exp2},
magnetic-field dependence of the thermal conductivity\cite{heat1,heat2}, 
and soundwave anomalies\cite{sound}.

Theoretical investigation was stimulated by Miyahara and Ueda\cite{miyahara1}
who pointed out the locality of the excited states in the SS model.
Owing to the expectation that this locality should account for the
observed magnetization plateaus, many calculations based on the
perturbative
methods\cite{Mueller,Weihong,momoi1,totsuka1,fukumoto,Knetter} 
have been carried out.  
Although their results qualitatively agree with 
the experimental data\cite{raman,infra,ESR,NMR},
the expansion parameter, estimated to be $0.6-0.7 $ for 
$\rm{ SrCu_2(BO_3)_2
}$, seems too large to ensure the validity of the perturbation.
Therefore non-perturbative approaches are desired in order to confirm 
discussions on the excited states.

In this paper we show numerical results obtained by a variational 
method, which we have recently developed and named the operator 
variational (OV) method\cite{ov1,ov2}, 
together with the restructuring technique\cite{munes,ehm}.
Our purpose is to study the lowest excited state of the SS model 
on a square lattice sketched in Fig.~1.
Using the OV method we can perform systematic calculations without
damaging the quantum symmetries of the system. The trial states,
on the other hand, should be given by inspection. 
We examine some of the one-triplet states and 
the two-triplet states with the total spin $S = 0$, 1 and 2, 
paying special attentions to the symmetries of the system's Hamiltonian 
on lattices equally sized in the $x$ and the $y$ directions.
We see that the value of the lowest energy, 
which provides an upper bound of the true value, 
strongly depends on the choice of the trial state. 
We therefore employ various trial states and adopt the one whose energy
 is the lowest among them. 

Our results indicate several interesting features on the dispersion 
relations and on the phase transition of the model. They are in good
accordance with the perturbative results if 
the inter-dimer coupling is weak, but discrepancies are observed 
when we move on toward the transition point. 
It is likely that the helical order phase\cite{koga1, helical} exists.
Trying a comparison with experimental data on 
$\rm{ SrCu_2(BO_3)_2 }$ we obtain several encouraging results.

In section 2 we introduce the SS model and discuss its symmetry used to
distinguish the states. Section 3 is to give a brief description on the 
OV method and to comment on its application to the SS model.
Numerical results on $8^2$ and $12^2$ lattices are presented in section 4 and
the final section is devoted to summary and discussions.
In appendices A-E we show the trial states employed in our numerical study.

\section{Model and Symmetry}

In the Shastry-Sutherland model a spin is located at every site of the
lattice schematically shown in Fig.~1, whose coordinates are denoted by
either $$ (2n_x a \pm \frac{d}{2} , 2n_y a \pm \frac{d}{2} ) \ \ \
 (A  \ {\rm sublattice}) \ \ \  {\rm or} \ \ \
((2n_x+1) a \pm \frac{d}{2} , (2n_y+1) a \mp \frac{d}{2} ) \ \ \ 
(B  \ {\rm sublattice}) $$
with $ n_x, n_y =0,1,\cdots ,N_{eff}-1$, 
where $N_{eff}$ is the number of dimers, 
$2a$ is the unit distance between dimers
and $d$ is the distance between two spins of a dimer. 
The total number of spins $N$ therefore equals to $4N_{eff}^2$.

The Hamiltonian of this model is given by 
\begin{eqnarray}
\hat{H} & = & \frac{1}{4} 
J \sum_{n_x,n_y=0}^{N_{eff}-1}  \{  h_a(n_x,n_y)+h_b(n_x,n_y) \}
\nonumber  \\
& + & \frac{1}{4} J'\sum_{n_x,n_y=0}^{N_{eff}-1}
 \{ h_1(n_x,n_y)+h_2(n_x,n_y)+h_3(n_x,n_y)+h_4(n_x,n_y) \} , 
\end{eqnarray}
where, \mbox{\boldmath $\sigma$}$(x,y)$ being the Pauli matrix at the
location $(x,y)$, 
\begin{eqnarray}
h_a(n_x,n_y) & \equiv & \mbox{\boldmath $\sigma$}(2n_x a + \frac{d}{2} , 2n_y a+ \frac{d}{2} ) 
\cdot  \mbox{\boldmath $\sigma$}(2n_x a - \frac{d}{2} , 2n_y a- \frac{d}{2} ), \\
h_b(n_x,n_y) & \equiv & \mbox{\boldmath $\sigma$}((2n_x+1)a + \frac{d}{2} , (2n_y+1) a- 
\frac{d}{2} ) \cdot 
 \mbox{\boldmath $\sigma$}((2n_x+1) a - \frac{d}{2} , (2n_y+1) a+ \frac{d}{2} ), \\
h_1(n_x,n_y) & \equiv & \mbox{\boldmath $\sigma$}(2n_x a + \frac{d}{2} , 2n_y a+ \frac{d}{2} ) 
\cdot 
\nonumber \\
    & &  \{ \mbox{\boldmath $\sigma$}((2n_x+1) a + \frac{d}{2} , (2n_y+1) a- \frac{d}{2} )
         + \mbox{\boldmath $\sigma$}((2n_x+1) a - \frac{d}{2} , (2n_y+1) a+ \frac{d}{2} ) \}, \\
h_2(n_x,n_y) & \equiv & \mbox{\boldmath $\sigma$}(2n_x a - \frac{d}{2} , 2n_y a- \frac{d}{2} ) 
\cdot  \nonumber \\
     & &   \{ \mbox{\boldmath $\sigma$}((2n_x-1) a + \frac{d}{2} , (2n_y-1) a- \frac{d}{2} )
          + \mbox{\boldmath $\sigma$}((2n_x-1) a - \frac{d}{2} , (2n_y-1) a+ \frac{d}{2} ) \}, \\
h_3(n_x,n_y) & \equiv &  \mbox{\boldmath $\sigma$}((2n_x+1) a + \frac{d}{2} , (2n_y+1) a- 
\frac{d}{2} )\cdot \nonumber \\
    & &   \{ \mbox{\boldmath $\sigma$}((2n_x+2) a + \frac{d}{2} , 2n_y a+ \frac{d}{2} ) +
            \mbox{\boldmath $\sigma$}((2n_x+2) a - \frac{d}{2} , 2n_y a- \frac{d}{2} ) \}, \\
h_4(n_x,n_y) & \equiv & \mbox{\boldmath $\sigma$}((2n_x+1) a - \frac{d}{2} , (2n_y+1) a+ 
\frac{d}{2} )\cdot  \nonumber \\
     & &   \{ \mbox{\boldmath $\sigma$}(2n_x a + \frac{d}{2} , (2n_y+2) a+ \frac{d}{2} ) +
             \mbox{\boldmath $\sigma$}(2n_x a - \frac{d}{2} , (2n_y+2) a- \frac{d}{2} ) \}. 
\end{eqnarray}
Periodic boundary conditions are assumed in both $x$ and $y$ directions.

Symmetric operates to commute with the Hamiltonian are 
$T_x$ (translation in the $x$ direction),  
$T_y$ (translation in the $y$ direction), 
$U$ ($A$-$B$ translation), 
$V_x$ ($A$-$B$ translation in the $x$ direction),  
$V_y$ ($A$-$B$ translation in the $y$ direction) and 
$I_\pm$ ($x$-$y$ reflections), which translate  \mbox{\boldmath
$\sigma$}$(x,y)$ as follows.
$$
T_x \mbox{\boldmath $\sigma$}(x,y)T_x^\dagger = \mbox{\boldmath $\sigma$}(x-2a,y), \ \ \
T_y \mbox{\boldmath $\sigma$}(x,y)T_y^\dagger = \mbox{\boldmath $\sigma$}(x,y-2a), $$
$$ 
U \mbox{\boldmath $\sigma$}(x,y)U^\dagger = \mbox{\boldmath $\sigma$}(
-y-a,x+a), $$ $$
V_x \mbox{\boldmath $\sigma$}(x,y)V_x^\dagger = \mbox{\boldmath $\sigma$}( -x-a,y+a), \ \ \
V_y \mbox{\boldmath $\sigma$}(x,y)V_y^\dagger = \mbox{\boldmath
$\sigma$}( x+a,-y-a), $$ $$
 I_+ \mbox{\boldmath $\sigma$}(x,y)I_+^\dagger = \mbox{\boldmath $\sigma$}( y,x), \ \ \
I_- \mbox{\boldmath $\sigma$}(x,y)I_-^\dagger = \mbox{\boldmath $\sigma$}(-y,-x). $$
Note that 
$$ U^4= \hat{1}, \ \ \  I_+^2= \hat{1}, \ \ \  I_-^2= \hat{1}, $$
where $\hat 1$ denotes the unit operator, and  
$$  U = I_+ V_y= I_- V_x .$$

\section{OV method}

In this section we make a brief description of the OV method we proposed
in ref.19, where we showed that this method enables us to 
calculate the energy  for large
quantum systems with less computer memory resources compared with 
the Lanczos method.

In the OV method, assuming that the Hamiltonian $\hat H$ is the sum of 
$N_g$ partial Hamiltonians $\hat g_i$, 
$$\hat H = \sum_{i=1}^{N_{g}} \hat g_{i} ,$$
we systematically generate candidate operators
$$ \hat O (k_1 \equiv 0 , k_2, \cdots, k_L) \equiv \sum_{i=1}^{N_{g}}
\hat g_{i+k_1} \hat g_{i+k_2} \cdots \hat g_{i+k_L} \ \ \ 
(L=1,2, \cdots, L_{max}),$$
with some integer $L_{max}$ and all possible values of integers $k_2$, 
$k_3$, $\cdots$, $k_L$.
Then we select a set of operators 
$ \{ \hat O_i \}=\{ \hat O_0 \equiv I (Identity) , \hat O_1, \cdots, 
\hat O_{n-1} \} $
from those candidates so that   
$\hat O_0 \mid \Psi \rangle$, $\  \hat O_1 \mid \Psi \rangle, \cdots,\hat
O_{n-1} \mid \Psi \rangle$ includes all linearly independent states 
for the given $L_{max}$ and $\mid \Psi \rangle$. 
Thus we obtain an approximate basis to calculate the Hamiltonian matrix
elements. Once the basis is determined, the
orthogonalization and the diagonalization are easily carried out 
by conventional methods.

Let us then concentrate our attention to trial states, 
the highly model-dependent component of the method.
We study the dispersion in two cases. One is the case
$p_y=p_x$, where we employ those states which have $even/odd$ parity 
in the $x$-$y$ reflection $I_+$. They are one-triplet states with 
the total spin $S=1$, 
two-nearest-neighboring-triplet states with $S = 0$, 1, 2 (abbreviated
as $nn$-triplet states hereafter) and 
two-next-nearest-neighboring-triplet states ($nnn$-triplet states 
hereafter) with $S = 0$, 1, 2.  
In the other case $p_y=0$, 
the trial states are eigenstates of $V_y$, the
operator to represent $A$-$B$ translation in the $y$ direction. 
Details of these trial states are given in appendices. 

\section{Results}

Now we present our results obtained on $8 \times 8$ and $12 \times
12$ lattices ($p_x$ or $p_y$ being therefore 0, $\frac{1}{3}\pi$, 
$\frac{1}{2}\pi$, $\frac{2}{3}\pi$ and ${\pi}$) with $L_{max}=3$.
We did not find any size effect on the results with the same ($p_x$,
$p_y$), where $p_x(p_y) = 0$ or $\pi$, calculated for these two lattice
sizes.\cite{footnote1,footnote2}

In Fig. 2 we plot, as a function of the coupling ratio $J'/J$, 
the difference between the energy of the singlet-dimer state 
$E_0$ and the lowest energy  $E$   
for the momentum \mbox{\boldmath $p$}=$(0,0)$,  
the total spin $S=j \ (j= 0, 1)$ and the $even$ or $odd$ $I_+$ parity. 
We determine $E$ with $S=1$  
as the minimum value among the energy  calculated from 
the one-triplet trial states, the $nn$-triplet trial states and  
the $nnn$-triplet trial states 
which are explicitly described in the Appendices D and E. Let us
introduce notations $E_{min}^{one}$, $E_{min}^{nn}$ and $E_{min}^{nnn}$
to represent the minimum energy  obtained from the one-, the $nn$-
and the $nnn$-triplet trial states, respectively.
Then the above is stated by 
$E \equiv min \{ E_{min}^{one}, E_{min}^{nn}, E_{min}^{nnn} \}$.
As for the $S=0$ states 
we compare the $nn$- and the $nnn$-triplet trial 
states, namely $E \equiv min \{ E_{min}^{nn}, E_{min}^{nnn} \}$.

When \mbox{\boldmath $p$}=$(0,0)$ it turned out for $S=1$ that 
$E = E_{min}^{nn}$ 
and $E$ is doubly degenerate, being independent on the $I_+$ parity.
Our results on the energy difference $E - E_0$, 
which correspond to the spin gap for this 
value of $S$, nicely agree with the perturbative expansion up to the 
fifth order of $J'/J$ (the dotted line in the figure) 
presented by Fukumoto\cite{fukumoto}. 
Especially for small
values of $J'/J$, say for $J'/J \le 0.5$, the agreement is excellent. 
When $S=0$, $E$ is obtained from the $nn$-triplet trial states for the
$odd$ $I_+$ parity in the whole range of the examined coupling
ratio, while for the $even$ $I_+$ parity $E=E_{min}^{nnn}$ 
if $J'/J \ge 0.55$.     
Both $I_+$ parity results are also compatible 
with the ones based on the perturbation
theory\cite{totsuka1},\cite{Knetter}. 

A remarkable feature in Fig.~2 is a crossover between $E$ with $S=0$, 
$odd$ $I_+$ parity and $E$ with $S=1$. We see that the former, which is 
much larger than the latter when $J'/J$ is 0.45, decreases more rapidly 
to reach $E_0$ faster than the latter when the coupling ratio grows. 
This observation suggests the existence\cite{footnote3}
of the helical order phase\cite{koga1,helical}  
between $r_1 < J'/J < r_2$, where $r_1$ ($r_2$)  
is the value of the coupling ratio where $E$ with $S=0$, $odd$ 
$I_+$ parity (with $S=1$, $even$ and $odd$ $I_+$ parity) 
becomes equal to $E_0$.
There are two reasons to believe that this crossover is not an artifact 
due to the approximations in the OV method but a real property of the 
SS model.   
One is that the exact calculations on small lattices (16 and 24 sites) also  
indicate this crossover. Another reason is that the $S=1$ results would
be less affected by the higher order corrections than the $S=0$ ones 
because they are known to be more reliable  
in the coupling region around the crossover.
Since the OV method is a kind of variational method, this means that  
we would observe more decrease of $E$ in the $S=0$ case 
when $L_{max}$ is increased. The
crossover is thus expected to survive in more precise estimations. 
Quantitatively, the crossover point and the values 
of $r_1$ and $r_2$, which are $\simeq 0.65$, $\simeq 0.75$ and 
$ \simeq 0.8$ in the present study, might change for the smaller 
in further numerical work.

Next we study how $E$ depends on the momentum $(p,p)$ or $(p,0)$.  
In Figs. 3-5 we show dispersions of the energy difference 
calculated for \mbox{\boldmath $p$}=$(p,p)$, $S=j \ (j= 0, 1, 2)$
and \mbox{\boldmath $p$}=$(p,0)$, $S=1$ 
with the coupling ratios 
$J'/J = 0.5 $ (Fig.~3), 0.6 (Fig.~4) and 0.65 (Fig.~5). 
Here we examine the one- and the $nn$-triplet trial 
states for $S=1$ and \mbox{\boldmath $p$}=$(p,0)$ , while for 
\mbox{\boldmath $p$}=$(p,p)$ we calculate 
$E \equiv min \{ E_{min}^{nn}, E_{min}^{nnn} \}$   
($E \equiv min \{ E_{min}^{one}, E_{min}^{nn}, E_{min}^{nnn} \}$) when
$S=0, 2$ ($S=1$).
  
Let us first direct our attention to the dispersions from the 
$S=1$ trial states. 
When \mbox{\boldmath $p$}=$(p,p)$ we observe, at each coupling ratio 
stated above, that the lowest energy  with even $I_+$ parity 
is $E_{min}^{nn}$ for all values of $p$. 
With odd $I_+$ parity, on the other hand, we observe that 
$E=E_{min}^{one}$ for $0 < p < \pi$. 
As was stated above, the lowest energy  for $p=0$ is
irrelevant to the $I_+$ parity so that the system has doubly
degenerate $E (=E_{min}^{nn}) $. This degeneracy is also found at $p=\pi$.
In another case \mbox{\boldmath $p$}=$(p,0)$, we find 
that $E_{min}^{nn}$ is slightly smaller than  
$E_{min}^{one}$ when $p=\pi$ while $E_{min}^{nn} > E_{min}^{one}$ for all
other values of $p$.
The $S=1$ results in the figures indicate that the 
$even$ $I_+$ state is almost dispersionless. Its value changes only 
$\sim$ 3~\% even when the value of $J'/J$ is as large as 0.65.
We also see that the energy with the $odd$ $I_+$ state 
depends very weakly on $p$. 
Similar tendency is observed for the \mbox{\boldmath $p$}=$(p,0)$ state, too. 
This feature is in marked contrast to the reported perturbative work, 
where the dispersion is flat only for small values of the coupling
ratio. For example, the dispersion by Weihong $et$ $al.$\cite{Weihong}
shows rapidly growing fluctuations near $J'/J \sim 0.65$. 
  
Turning to results for $S=0$ state we see strong momentum dependence of
$E$ both for $even$ and $odd$ $I_+$ parity, which is the very 
impressive characteristic of this state. 
Here $E=E_{min}^{nn}$ always holds if the $I_+$ parity is $odd$.
For the $even$ $I_+$ parity, however, 
it depends on the values of $J'/J$ and $p$
whether $E_{min}^{nn} < E_{min}^{nnn}$ or not.
When the coupling ratio is 0.5 it turned out that $E=E_{min}^{nn}$ 
whatever $p$ is.
On the other hand, we observe $E_{min}^{nnn}$ underlies $E_{min}^{nn}$ 
for $J'/J= 0.6$, $p=0$ and for $J'/J= 0.65$, $p=0$ or $p = \pi /3$.
These dispersions are in qualitative agreement with those given 
by Totsuka $et$ $al.$\cite{totsuka1} for $J'/J$=0.5 
and by Fukumoto\cite{fukumoto} for $J'/J$=0.55, 
although their values are smaller than ours.  
 
Finally we comment on the $S=2$ case, where the relation  
$E_{min}^{nnn} < E_{min}^{nn}$ is always observed\cite{footnote4}.
Both the $even$ and the $odd$ $I_+$ parity dispersions are weakly
dependent on the momentum in accordance with statements in refs. 12 and 13.
Our results did not confirm, however, 
their claim that the energy difference is smaller than the twice of the 
spin gap so that the $S=2$ state is a stable bound state composed of the
lowest excited state. What we read from Figs. 3-5 is that $E-E_0$ for
$S=2$ is located slightly above the twice of that for $S=1$. 

Before proceeding to the next section we would like to make a comparison
between our results and the experimental data for SrCu$_2$(BO$_3$)$_2$.
The energy of the lowest excited state with $S=1$, namely the spin
gap, can be determined in a very reliable manner both in
experiments and in numerical works. The spin gap for this matter is
established to be 3.0 meV (24.2 cm$^{-1}$, 34.7 K) through various 
experiments\cite{exp2,infra,ESR}.  
Another reliable quantity shared by experimental and theoretical 
investigations would be the energy of the lowest excited state 
with $S=0$, which is 
reported to be 30 cm$^{-1}$ (3.7 meV) in the Raman scattering experiment
\cite{raman}. According to Knetter $et$ $al.$\cite{Knetter},
the observed energy is that of the SD=$-$ state (the $odd$ $U$ parity 
state in our terminology) because the SD=$+$ state, which has the lower
energy, is forbidden in this experiment. Among our trial states with 
$S=0$ and \mbox{\boldmath $p$}=$(0,0)$ shown in the Appendix D, the $odd$
$U(=I_- V_x)$ ones are $\mid \Psi_{nn,0,(0,0),a'} \rangle$, 
$\mid \Psi_{nnn,0,(0,0),b'}\rangle$ and 
 $\mid \Psi_{nnn,0,(0,0),c'}\rangle$ while 
$\mid \Psi_{nn,0,(0,0),b'}\rangle$, $\mid \Psi_{nnn,0,(0,0),a'}\rangle$ and 
$\mid \Psi_{nnn,0,(0,0),d'}\rangle$ have the $even$ $U$ parity.
What we observe are that the lowest (the second lowest) energy
 is $E_{nn,0,(0,0),a'}$ ($E_{nn,0,(0,0),b'}$) if the coupling
ratio is less than or equal to 0.8 (0.7).
Thus, although we plot $E_{nn,0,(0,0),a'}$ in Fig.~2, we should
examine $E_{nn,0,(0,0),b'}$ here. Requesting that the ratio 
$(E_{nn,0,(0,0),b'}-E_0)/(E_{nn,1,(0,0),c'}-E_0)$ should be equal to 
the experimental value 3.7/3.0 = 1.23, where $E_{nn,1,(0,0),c'}$ is 
the lowest energy  with $S=1$,  
we fix $J'/J$ for SrCu$_2$(BO$_3$)$_2$ 
is 0.65 and $J = 87$K. These values are very close to those estimated by
in ref.29, which are 0.635 and 85K, respectively. 

Table~1 lists up our results for $S \le 1$ at 
$J'/J=0.65$ which are the lowest energy  with different
symmetries. From Table~I we can further estimate, using 
the value $J = 87$K, that the energy difference  
of the $odd$ $U$, $even$ $I_+$ state (of the $odd$ $I_+ I_-$ 
state) is about 6.8 meV (7.0 meV). These might
correspond to the $S=0$ state observed in the infrared 
experiment\cite{infra}, whose energy is 52 cm$^{-1}$ (6.4 meV), 
or the one in the Raman scattering\cite{raman}, whose energy is 
56 cm$^{-1}$ (6.9 meV). 
We did not find, however, 
any $S=1$ state to explain the second excited
state observed in the experiments\cite{exp2,ESR} 
with the energy 54.7K (4.71 meV).  The reason would be that the OV
method is a kind of variational approach where the higher
excitations for each set of quantum numbers are difficult to calculate. 

\section{Summary and discussions}

In this paper we made a report of the excited states of the SS model. 
We carried out numerical investigations using the OV method, 
a kind of variational method we have developed\cite{ov1,ov2}. 
Our results show that the dispersion of the lowest excited state 
with the total spin $S=1$ is quite small, 
which indicates the locality of the state, 
even when the coupling ratio $J'/J$ is as large as 0.65.
It provides a contrast to the perturbative results, where the small dispersion
is observed only for small $J'/J$\cite{Weihong}.
Another noticeable feature in our observations is that, when we start
from $J'/J=0.45$ and increase this ratio keeping the momentum 
\mbox{\boldmath $p$}=\mbox{\boldmath $0$}, 
the energy of the $S=0$ state decreases faster 
than the energy of the $S=1$ state to form a crossover near $J'/J = 0.65$.
This crossover supports an existence of the helical order phase 
suggested in refs. 23 and 24. 
The upper bound of $J'/J$ for the phase transition point between the 
orthogonal dimer phase and the helical order phase is 0.75 in this study.
We also successfully estimate the values of $J'/J$ and $J$ from our
results and the experimental data, $J'/J= 0.65$ and $J=87$K for 
SrCu$_2$(BO$_3$)$_2$, which are compatible with
the preceding studies\cite{Knetter,miyahara2}.

Since it is so far difficult to perform the next order ($L_{max}=4$)
calculations, which would be the
best way to figure out the errors in our results, let us here try to 
discuss the reliability of our data (labeled by OV hereafter) 
by comparing some of them with the perturbative results   
\cite{totsuka1,fukumoto,Knetter}
(PB12,PB13,PB14),
with the values from the perturbative expansion on the spin gap up to the fifth
order of $J'/J$\cite{fukumoto} (PBF) 
and with the results from the exact diagonalization of 24 or 16 site  
cluster (ED24, ED16). For the lowest $S=1$ state, we have 0.679 (OV), 0.685
(PBF), 0.6779 (ED24) and 0.6797 (ED16) at $J'/J = 0.5$ while they are 
0.402 (OV), 0.436 (PBF), 0.3676 (ED24) and 0.3836 (ED16) when $J'/J = 0.65$. 
In the similar analysis on the lowest excited state with $S=0$, we
obtain 0.93 (OV), 0.80 (PB12), 
0.90 (PB13, PB14), 0.858 (ED24) and 0.817
(ED16) at $J'/J = 0.5$ while they are 0.59 (OV), 0.55
(PB13), 0.50 (PB14), 0.432 (ED24) and 0.386
(ED16) when $J'/J = 0.6$. These results suggest that our data are quite
reliable when $J'/J \sim 0.5$ and for larger values of the coupling
ratio they give us  reasonable upper 
bounds of the energy difference, especially for the spin gap. 

We did not observe any instability on the $S=1$ two triplet states
for large $J'/J$ in contrast to the report  in refs.13 and 14.
 Because this is an important issue in
connection with the existence of the helical order phase, further study
would be necessary before we draw a definite conclusion.
Another physically intriguing phenomenon is the magnetization plateau of    
SrCu$_2$(BO$_3$)$_2$. It is so far an open question to explain the 
experimentally observed 1/8 plateau from a theoretical point of 
view\cite{momoi1,fukumoto2}.
We hope our method is helpful to solve this problem.       

\appendix

\section{Ground state}
It is known that the exact ground state $ \mid \Psi _0  \rangle$
for $J'/J \leq 0.68 $ is the product of all singlet dimers,
\begin{eqnarray*}
 \mid \Psi _0  \rangle = \prod_{n_x,n_y=0}^{N_{eff}-1}
 \mid singlet(2n_x,2n_y) \rangle \cdot \mid singlet(2n_x+1,2n_y+1) \rangle ,
\end{eqnarray*}
with
$$ \mid  singlet(2n_x,2n_y) \rangle  \equiv $$
$$ \frac{1}{\sqrt{2}}
\{  \mid \uparrow(2n_x a + \frac{d}{2} , 2n_y a+ \frac{d}{2} )
\downarrow(2n_x a - \frac{d}{2} , 2n_y a- \frac{d}{2} ) \rangle $$
$$ -  \mid \downarrow(2n_x a + \frac{d}{2} , 2n_y a+ \frac{d}{2} )
\uparrow(2n_x a - \frac{d}{2} , 2n_y a- \frac{d}{2} ) \rangle \} , $$$$ \mid singlet(2n_x+1,2n_y+1) \rangle  \equiv   $$ 
$$ \frac{1}{\sqrt{2}}
\{\mid\uparrow((2n_x+1) a + \frac{d}{2} , (2n_y+1) a- \frac{d}{2} )
\downarrow((2n_x+1) a - \frac{d}{2} , (2n_y+1) a+ \frac{d}{2} ) \rangle  $$
$$-\mid\downarrow((2n_x+1) a + \frac{d}{2} , (2n_y+1) a- \frac{d}{2} )
\uparrow((2n_x+1) a -\frac{d}{2},(2n_y+1) a+ \frac{d}{2} ) \rangle \} .$$

\section{Operators which create the triplet dimer states}
We introduce the operators $\hat T_j \ (j= 0, \ \pm 1) \ $ which operate
the dimer singlet state and create the triplet one,  
\begin{eqnarray*} 
& & \hat{T}_j(2n_x,2n_y) \mid signlet(2n_x,2n_y)  \rangle = \mid
S_z=j(2n_x,2n_y)  \rangle  , \\
& & \hat{T}_j(2n_x+1,2n_y+1) \mid signlet(2n_x+1,2n_y+1)  \rangle = \mid
S_z=j(2n_x+1,2n_y+1)  \rangle ,  
\end{eqnarray*}
where 
\begin{eqnarray*}
& & \hat{T}_1(2n_x,2n_y) \equiv
 \frac{1}{\sqrt{2}}
\{- \sigma^+(2n_x a + \frac{d}{2} , 2n_y a+ \frac{d}{2} ) +
      \sigma^+(2n_x a - \frac{d}{2} , 2n_y a- \frac{d}{2} ) \}, \\
& & \hat{T}_1(2n_x+1,2n_y+1) \equiv \\
& & \frac{1}{\sqrt{2}}
 \{- \sigma^+((2n_x+1) a + \frac{d}{2} , (2n_y +1)a- \frac{d}{2} ) +
      \sigma^+((2n_x+1) a - \frac{d}{2} , (2n_y+1) a+ \frac{d}{2} ) \}  , 
\end{eqnarray*}
\begin{eqnarray*}
 & & \hat{T}_{-1}(2n_x,2n_y) \equiv
- \frac{1}{\sqrt{2}}
\{- \sigma^-(2n_x a + \frac{d}{2} , 2n_y a+ \frac{d}{2} ) +
      \sigma^-(2n_x a - \frac{d}{2} , 2n_y a- \frac{d}{2} ) \} , \\
& & \hat{T}_{-1}(2n_x+1,2n_y+1) \equiv \\ 
& & -\frac{1}{\sqrt{2}}
 \{- \sigma^-((2n_x+1) a + \frac{d}{2} , (2n_y +1)a- \frac{d}{2} ) +
      \sigma^-((2n_x+1) a - \frac{d}{2} ,(2n_y+1) a+ \frac{d}{2} ) \} ,
\end{eqnarray*}
\begin{eqnarray*}
& & \hat{T}_{0}(2n_x,2n_y) \equiv
- \frac{1}{2}
\{ - \sigma^z(2n_x a + \frac{d}{2} , 2n_y a+ \frac{d}{2} ) +
      \sigma^z(2n_x a - \frac{d}{2} , 2n_y a- \frac{d}{2} ) \} , \\
& & \hat{T}_{0}(2n_x+1,2n_y+1) \equiv \\ 
& & -\frac{1}{2}
 \{ - \sigma^z((2n_x+1) a + \frac{d}{2} , (2n_y +1)a- \frac{d}{2} ) +
      \sigma^z((2n_x+1) a - \frac{d}{2} , (2n_y+1) a+ \frac{d}{2} ) \}  .
\end{eqnarray*}

\section{States in the momentum space}
We define the states in the momentum space as follows.
\begin{eqnarray*}
 \mid (p_x,p_y) A \rangle & \equiv & \sum_{n_x,n_y} \exp( ip_x n_x +ip_y n_y)
   \hat{T}_1(2n_x,2n_y)\mid\Psi_0  \rangle ,\\
 \mid (p_x,p_y) B \rangle & \equiv & \sum_{n_x,n_y} \exp( ip_x n_x +ip_y n_y)
   \hat{T}_1(2n_x+1,2n_y+1)\mid\Psi_0  \rangle  ,
\end{eqnarray*}
\begin{eqnarray*}
 \mid (p_x,p_y) AR2 \rangle & \equiv & \sum_{n_x,n_y} \exp( ip_x n_x +ip_y n_y)
   \hat{T}_1(2n_x,2n_y) \hat{T}_1(2n_x+1,2n_y+1)\mid\Psi_0  \rangle , \\
 \mid (p_x,p_y) AL2 \rangle & \equiv & \sum_{n_x,n_y} \exp( ip_x n_x +ip_y n_y)
   \hat{T}_1(2n_x,2n_y) \hat{T}_1(2n_x-1,2n_y-1)\mid\Psi_0  \rangle  ,\\
 \mid (p_x,p_y) BR2 \rangle & \equiv & \sum_{n_x,n_y} \exp( ip_x n_x +ip_y n_y)
   \hat{T}_1(2n_x+1,2n_y+1) \hat{T}_1(2n_x+2,2n_y)\mid\Psi_0  \rangle , \\
 \mid (p_x,p_y) BL2 \rangle & \equiv & \sum_{n_x,n_y} \exp( ip_x n_x +ip_y n_y)
   \hat{T}_1(2n_x+1,2n_y+1) \hat{T}_1(2n_x,2n_y+2)\mid\Psi_0  \rangle ,
\end{eqnarray*}
\begin{eqnarray*} 
& & \mid (p_x,p_y) AR1 \rangle  \equiv  \sum_{n_x,n_y} \exp( ip_x n_x +ip_y n_y)
  \cdot \\
& & \lbrace  \hat{T}_1(2n_x,2n_y) \hat{T}_0(2n_x+1,2n_y+1)
- \hat{T}_0(2n_x,2n_y) \hat{T}_1(2n_x+1,2n_y+1) \rbrace \mid\Psi_0  \rangle ,\\
& &
\mid (p_x,p_y) AL1 \rangle  \equiv  \sum_{n_x,n_y} \exp( ip_x n_x +ip_y n_y)
\cdot  \\
& & \lbrace  \hat{T}_1(2n_x,2n_y) \hat{T}_0(2n_x-1,2n_y-1)
- \hat{T}_0(2n_x,2n_y) \hat{T}_1(2n_x-1,2n_y-1) \rbrace \mid\Psi_0  \rangle ,\\
& & \mid (p_x,p_y) BR1 \rangle \equiv \sum_{n_x,n_y} \exp( ip_x n_x +ip_y n_y)
\cdot  \\
& & \lbrace  \hat{T}_1(2n_x+1,2n_y+1) \hat{T}_0(2n_x+2,2n_y)
- \hat{T}_0(2n_x+1,2n_y+1) \hat{T}_1(2n_x+2,2n_y) \rbrace \mid\Psi_0 \rangle ,\\
& & \mid (p_x,p_y) BL1 \rangle \equiv \sum_{n_x,n_y} \exp( ip_x n_x +ip_y n_y)
\cdot  \\
& & \lbrace  \hat{T}_1(2n_x+1,2n_y+1) \hat{T}_0(2n_x,2n_y+2)
- \hat{T}_0(2n_x+1,2n_y+1) \hat{T}_1(2n_x,2n_y+2) \rbrace \mid\Psi_0  \rangle ,
\end{eqnarray*}
\begin{eqnarray*} 
& &\mid (p_x,p_y) AR0 \rangle  \equiv  \sum_{n_x,n_y} \exp( ip_x n_x +ip_y n_y)
 \cdot \\
& & \lbrace  \hat{T}_1(2n_x,2n_y) \hat{T}_{-1}(2n_x+1,2n_y+1)
+ \hat{T}_{-1}(2n_x,2n_y) \hat{T}_1(2n_x+1,2n_y+1)  \\
& &- \hat{T}_0(2n_x,2n_y) \hat{T}_0(2n_x+1,2n_y+1)\rbrace \mid\Psi_0 \rangle ,\\
& &\mid (p_x,p_y) AL0 \rangle  \equiv  \sum_{n_x,n_y} \exp( ip_x n_x +ip_y n_y)
\cdot  \\
& & \lbrace  \hat{T}_1(2n_x,2n_y) \hat{T}_{-1}(2n_x-1,2n_y-1)
+ \hat{T}_{-1}(2n_x,2n_y) \hat{T}_1(2n_x-1,2n_y-1)  \\
& &- \hat{T}_0(2n_x,2n_y) \hat{T}_0(2n_x-1,2n_y-1) \rbrace\mid\Psi_0 \rangle ,\\
& & \mid (p_x,p_y) BR0 \rangle  \equiv  \sum_{n_x,n_y} \exp( ip_x n_x +ip_y n_y)
\cdot  \\
& & \lbrace  \hat{T}_1(2n_x+1,2n_y+1) \hat{T}_{-1}(2n_x+2,2n_y)
+ \hat{T}_{-1}(2n_x+1,2n_y+1) \hat{T}_1(2n_x+2,2n_y)  \\
& &- \hat{T}_0(2n_x+1,2n_y+1)\hat{T}_0(2n_x+2,2n_y)\rbrace \mid\Psi_0\rangle ,\\
& & \mid (p_x,p_y) BL0 \rangle  \equiv  \sum_{n_x,n_y} \exp( ip_x n_x +ip_y n_y)
\cdot  \\
& & \lbrace  \hat{T}_1(2n_x+1,2n_y+1) \hat{T}_{-1}(2n_x,2n_y+2)
+ \hat{T}_{-1}(2n_x+1,2n_y+1) \hat{T}_1(2n_x,2n_y+2)\\
& &- \hat{T}_0(2n_x+1,2n_y+1)\hat{T}_0(2n_x,2n_y+2)\rbrace \mid\Psi_0 \rangle ,
\end{eqnarray*}
\begin{eqnarray*}
 \mid (p_x,p_y) Ax2 \rangle & \equiv & \sum_{n_x,n_y} \exp( ip_x n_x +ip_y n_y)
   \hat{T}_1(2n_x,2n_y) \hat{T}_1(2n_x+2,2n_y)\mid\Psi_0  \rangle , \\
 \mid (p_x,p_y) Ay2 \rangle & \equiv & \sum_{n_x,n_y} \exp( ip_x n_x +ip_y n_y)
   \hat{T}_1(2n_x,2n_y) \hat{T}_1(2n_x,2n_y+2)\mid\Psi_0  \rangle ,\\
 \mid (p_x,p_y) Bx2 \rangle & \equiv & \sum_{n_x,n_y} \exp( ip_x n_x +ip_y n_y)
   \hat{T}_1(2n_x+1,2n_y+1) \hat{T}_1(2n_x+3,2n_y+1)\mid\Psi_0  \rangle ,\\
 \mid (p_x,p_y) By2 \rangle & \equiv & \sum_{n_x,n_y} \exp( ip_x n_x +ip_y n_y)
   \hat{T}_1(2n_x+1,2n_y+1) \hat{T}_1(2n_x+1,2n_y+3)\mid\Psi_0  \rangle ,
\end{eqnarray*}
\begin{eqnarray*} 
& & \mid (p_x,p_y) Ax1 \rangle  \equiv  \sum_{n_x,n_y} \exp( ip_x n_x +ip_y n_y)
  \cdot \\
& & \lbrace  \hat{T}_1(2n_x,2n_y) \hat{T}_0(2n_x+2,2n_y)
- \hat{T}_0(2n_x,2n_y) \hat{T}_1(2n_x+2,2n_y) \rbrace \mid\Psi_0  \rangle ,\\
& &
\mid (p_x,p_y) Ay1 \rangle  \equiv  \sum_{n_x,n_y} \exp( ip_x n_x +ip_y n_y)
\cdot  \\
& & \lbrace  \hat{T}_1(2n_x,2n_y) \hat{T}_0(2n_x,2n_y+2)
- \hat{T}_0(2n_x,2n_y) \hat{T}_1(2n_x,2n_y+2) \rbrace \mid\Psi_0  \rangle ,\\
& & \mid (p_x,p_y) Bx1 \rangle \equiv \sum_{n_x,n_y} \exp( ip_x n_x +ip_y n_y)
\cdot  \\
& & \lbrace  \hat{T}_1(2n_x+1,2n_y+1) \hat{T}_0(2n_x+3,2n_y+1)
- \hat{T}_0(2n_x+1,2n_y+1) \hat{T}_1(2n_x+3,2n_y+1)
 \rbrace \mid\Psi_0  \rangle , \\
& & \mid (p_x,p_y) By1 \rangle \equiv \sum_{n_x,n_y} \exp( ip_x n_x +ip_y n_y)
\cdot  \\
& & \lbrace  \hat{T}_1(2n_x+1,2n_y+1) \hat{T}_0(2n_x+1,2n_y+3)
- \hat{T}_0(2n_x+1,2n_y+1) \hat{T}_1(2n_x+1,2n_y+3) \rbrace\mid\Psi_0 \rangle .
\end{eqnarray*}

\section{Trial states in the $p_y = p_x$ case}
Using the states defined in the previous section we construct trial states. 
We employ $S_z=S=j \ (j=0, 1, 2)$ states described below as the trial
states. It should be noted that some states are degenerate.
In order to point out the degeneracy we use the notation $E_m$ to
represent the lowest energy  obtained with the
trial state $ \mid \Psi_m \rangle$.

\subsection{one-triplet states ($S=1)$}
The trial states with one triplet are as follows.
\begin{eqnarray*}   
 \mid \Psi_{one,1,(p,p),a} \rangle & \equiv & \mid (p,p)A \rangle
 \ \ \ (I_+even) ,\\
 \mid \Psi_{one,1,(p,p),b} \rangle & \equiv & \mid (p,p)B \rangle
 \ \ \ (I_+odd ) .
\end{eqnarray*}  
Note that $E_{one,1,(p,p),a} = E_{one,1,(p,p),b}$. 

The following should also be kept in mind 
in order to make a comparison between
the numerical results and the experimental data.
$$   
 I_- \mid \Psi_{one,1,(0,0),a} \rangle =  - \mid \Psi_{one,1,(0,0),a}
 \rangle , \ \ \ 
 V_x \mid \Psi_{one,1,(0,0),a} \rangle =  - \mid \Psi_{one,1,(0,0),b}
 \rangle , $$
$$
 I_- \mid \Psi_{one,1,(0,0),b} \rangle =  + \mid \Psi_{one,1,(0,0),b}
 \rangle , \ \ \ 
V_x \mid \Psi_{one,1,(0,0),b} \rangle =  - \mid \Psi_{one,1,(0,0),a}\rangle .
$$ 

\subsection{two-nearest-neighboring-triplet states ($S=j, \ j=0, 1,
2$)}
The trial states with two nearest neighboring triplets
for $p \not= 0$, $p \not= \pi$ are  
\begin{eqnarray*}
\mid \Psi_{nn,j,(p,p),a} \rangle & \equiv & \mid (p,p)ARj 
\rangle \ \ \ (I_+odd) , \\ 
\mid \Psi_{nn,j,(p,p),b} \rangle & \equiv & \mid (p,p)ALj 
\rangle \ \ \ (I_+odd) , \\ 
\mid \Psi_{nn,j,(p,p),c} \rangle & \equiv & \mid (p,p)BRj
 \rangle + \mid (p,p)BLj \rangle \ \ \ (I_+odd) , \\ 
\mid \Psi_{nn,j,(p,p),d} \rangle & \equiv & \mid (p,p)BRj
 \rangle - \mid (p,p)BLj \rangle \ \ \ (I_+even) .
\end{eqnarray*}
For the momentum (0,0) they are
\begin{eqnarray*}
 \mid \Psi_{nn,j,(0,0),a'} \rangle & \equiv &  \mid(0,0)ARj  \rangle+
\mid(0,0)ALj  \rangle+\mid(0,0)BRj  \rangle+\mid(0,0)BLj  \rangle
\ \ \ (I_+odd) , \\
 \mid \Psi_{nn,j,(0,0),b'} \rangle & \equiv &   \mid(0,0)ARj  \rangle+
\mid(0,0)ALj  \rangle-\mid(0,0)BRj  \rangle-\mid(0,0)BLj  \rangle
 \ \ \ (I_+odd) , \\
 \mid \Psi_{nn,j,(0,0),c'} \rangle & \equiv &   \mid(0,0)ARj 
 \rangle- \mid(0,0)ALj  \rangle \ \ \ (I_+odd) , \\
 \mid \Psi_{nn,j,(0,0),d'} \rangle & \equiv &   \mid(0,0)BRj 
 \rangle- \mid(0,0)BLj  \rangle\ \ \ (I_+even) .
\end{eqnarray*}
Also for the momentum $(\pi,\pi)$ they are
\begin{eqnarray*}
 \mid \Psi_{nn,j,(\pi,\pi),a''} \rangle & \equiv &
 \mid(\pi,\pi)ARj  \rangle+ \mid(\pi,\pi)ALj  \rangle+
i\{\mid(\pi,\pi)BRj  \rangle+\mid(\pi,\pi)BLj  \rangle \} \ \ \ (I_+odd) , \\
 \mid  \Psi_{nn,j,(\pi,\pi),b''}\rangle & \equiv & \mid(\pi,\pi)ARj
\rangle+ \mid(\pi,\pi)ALj  \rangle-
i\{\mid(\pi,\pi)BRj  \rangle+\mid(\pi,\pi)BLj  \rangle\}\ \ \ (I_+odd) , \\
 \mid \Psi_{nn,j,(\pi,\pi),c''} \rangle & \equiv & \mid(\pi,\pi)ARj
\rangle- \mid(\pi,\pi)ALj  \rangle \ \ \ (I_+odd) , \\
 \mid \Psi_{nn,j,(\pi,\pi),d''} \rangle & \equiv & \mid(\pi,\pi)BRj
\rangle- \mid(\pi,\pi)BLj  \rangle \ \ \ (I_+even) .
\end{eqnarray*}

Note that
$E_{nn,j,(0,0),c'} =E_{nn,j,(0,0),d'}$ and  
$E_{nn,j,(\pi,\pi),c''} = E_{nn,j,(\pi,\pi),d''}$.

The following should also be kept in mind, too. 
$$ I_- \mid \Psi_{nn,j,(0,0),a'} \rangle =- \mid  \Psi_{nn,j,(0,0),a'} \rangle , \ \ \
V_x \mid  \Psi_{nn,j,(0,0),a'} \rangle =+ \mid \Psi_{nn,j,(0,0),a'} \rangle, $$
$$ I_- \mid  \Psi_{nn,j,(0,0),b'} \rangle =- \mid  \Psi_{nn,j,(0,0),b'} \rangle  , \ \ \
 V_x \mid  \Psi_{nn,j,(0,0),b'} \rangle = -\mid \Psi_{nn,j,(0,0),b'}\rangle, $$
$$ I_- \mid  \Psi_{nn,j,(0,0),c'} \rangle = +\mid   \Psi_{nn,j,(0,0),c'}\rangle  , \ \ \
 V_x \mid  \Psi_{nn,j,(0,0),c'} \rangle = -\mid\Psi_{nn,j,(0,0),d'} \rangle , $$
$$ I_- \mid  \Psi_{nn,j,(0,0),d'} \rangle = -\mid   \Psi_{nn,j,(0,0),d'}\rangle  , \ \ \
 V_x \mid  \Psi_{nn,j,(0,0),d'} \rangle = -\mid \Psi_{nn,j,(0,0),c'}\rangle . $$

\subsection{two-next-nearest-neighboring-triplet states ($S=j, \ j=0, 1, 2$)}
When $S=j'\ (j'=0,2)$ and $p=0$, 
 the trial states with two-next-nearest-neighboring triplets are
\begin{eqnarray*}
 \mid \Psi_{nnn,j',(0,0),a'} \rangle & \equiv &  \mid(0,0)Axj'  \rangle+
\mid(0,0)Ayj'  \rangle+\mid(0,0)Bxj'  \rangle+\mid(0,0)Byj'  \rangle
\ \ \ (I_+even) , \\
 \mid \Psi_{nnn,j',(0,0),b'} \rangle & \equiv &  \mid(0,0)Axj' \rangle+
\mid(0,0)Ayj'  \rangle-\mid(0,0)Bxj'  \rangle-\mid(0,0)Byj'  \rangle
\ \ \ (I_+even) , \\ 
 \mid \Psi_{nnn,j',(0,0),c'} \rangle & \equiv &  \mid(0,0)Axj'  \rangle-
\mid(0,0)Ayj'  \rangle+\mid(0,0)Bxj'  \rangle-\mid(0,0)Byj'  \rangle
\ \ \ (I_+odd) , \\
 \mid \Psi_{nnn,j',(0,0),d'} \rangle & \equiv &  \mid(0,0)Axj'  \rangle-
\mid(0,0)Ayj'  \rangle-\mid(0,0)Bxj'  \rangle+\mid(0,0)Byj' \rangle
\ \ \ (I_+odd)  .
\end{eqnarray*}
When $S=1$ and $p=\pi$ they are
\begin{eqnarray*}
 \mid \Psi_{nnn,1,(\pi,\pi),a''} \rangle & \equiv &  \mid(\pi,\pi)Ax1  \rangle+
\mid(\pi,\pi)Ay1  \rangle+i\{\mid(\pi,\pi)Bx1  \rangle+\mid(\pi,\pi)By1  \rangle\}
\ \ \ (I_+even) , \\
 \mid \Psi_{nnn,1,(\pi,\pi),b''} \rangle & \equiv &  \mid(\pi,\pi)Ax1  \rangle-
\mid(\pi,\pi)Ay1  \rangle-i\{\mid(\pi,\pi)Bx1  \rangle+\mid(\pi,\pi)By1  \rangle\}
\ \ \ (I_+odd) , \\ 
 \mid \Psi_{nnn,1,(\pi,\pi),c''} \rangle & \equiv &  \mid(\pi,\pi)Ax1  \rangle+
\mid(\pi,\pi)Ay1  \rangle+i\{\mid(\pi,\pi)Bx1  \rangle-\mid(\pi,\pi)By1  \rangle\}
\ \ \ (I_+even) , \\
 \mid \Psi_{nnn,1,(\pi,\pi),d''} \rangle & \equiv &  \mid(\pi,\pi)Ax1  \rangle-
\mid(\pi,\pi)Ay1  \rangle-i\{\mid(\pi,\pi)Bx1  \rangle-\mid(\pi,\pi)By1  \rangle\}
\ \ \ (I_+odd)  .
\end{eqnarray*}
Otherwise they are
\begin{eqnarray*}
\mid \Psi_{nnn,j,(p,p),a} \rangle & \equiv & \mid (p,p)Axj \rangle+\mid (p,p)Ayj \rangle
 \ \ \ (I_+even) , \\ 
\mid \Psi_{nnn,j,(p,p),b} \rangle & \equiv & \mid (p,p)Axj \rangle-\mid (p,p)Ayj \rangle
 \ \ \ (I_+odd) ,\\ 
\mid \Psi_{nnn,j,(p,p),c} \rangle & \equiv & \mid (p,p)Bxj \rangle+\mid (p,p)Byj \rangle
 \ \ \ (I_+even) , \\ 
\mid \Psi_{nnn,j,(p,p),d} \rangle & \equiv & \mid (p,p)Bxj \rangle-\mid (p,p)Byj \rangle
 \ \ \ (I_+odd) .
\end{eqnarray*}
Note that $E_{nnn,j',(\pi,\pi),a} = E_{nnn,j',(\pi,\pi),d}$, 
$E_{nnn,j',(\pi,\pi),b} = E_{nnn,j',(\pi,\pi),c}$, where $j'=0, 2$, and
$E_{nnn,1,(0,0),a} = E_{nnn,1,(0,0),d}$,  
$E_{nnn,1,(0,0),b} = E_{nnn,1,(0,0),c}$,
$E_{nnn,1,(\pi,\pi),a''} = E_{nnn,1,(\pi,\pi),c''}$ and 
$E_{nnn,1,(\pi,\pi),b''} = E_{nnn,1,(\pi,\pi),d''}$. 

The following should be kept in mind, too ( $j'=0,2$). 
$$ I_- \mid \Psi_{nnn,j',(0,0),a'} \rangle =+ \mid  \Psi_{nnn,j',(0,0),a'} \rangle , \ \ \
V_x \mid \Psi_{nnn,j',(0,0),a'}\rangle =+ \mid \Psi_{nnn,j',(0,0),a'}
\rangle . $$
$$ I_- \mid  \Psi_{nnn,j',(0,0),b'} \rangle =+ \mid  \Psi_{nnn,j',(0,0),b'} \rangle  , \ \ \
 V_x \mid  \Psi_{nnn,j',(0,0),b'} \rangle = - \mid  \Psi_{nnn,j',(0,0),b'}
 \rangle , $$
$$ I_- \mid  \Psi_{nnn,j',(0,0),c'} \rangle = -\mid   \Psi_{nnn,j',(0,0),c'}\rangle  , \ \ \
 V_x \mid  \Psi_{nnn,j',(0,0),c'} \rangle = +\mid  \Psi_{nnn,j',(0,0),c'}
 \rangle , $$
$$ I_- \mid  \Psi_{nnn,j',(0,0),d'} \rangle = -\mid   \Psi_{nnn,j',(0,0),d'}\rangle  , \ \ \
 V_x \mid  \Psi_{nnn,j',(0,0),d'} \rangle = -\mid  \Psi_{nnn,j',(0,0),d'}
 \rangle , $$
$$ I_- \mid \Psi_{nnn,1,(0,0),a} \rangle =- \mid  \Psi_{nnn,1,(0,0),a} \rangle , \ \ \
V_x \mid  \Psi_{nnn,1,(0,0),a} \rangle =- \mid \Psi_{nnn,1,(0,0),d} \rangle , $$
$$ I_- \mid \Psi_{nnn,1,(0,0),b} \rangle =+\mid  \Psi_{nnn,1,(0,0),b} \rangle , \ \ \
V_x \mid  \Psi_{nnn,1,(0,0),b} \rangle =- \mid  \Psi_{nnn,1,(0,0),c}
 \rangle , $$
$$ I_- \mid \Psi_{nnn,1,(0,0),c} \rangle =- \mid  \Psi_{nnn,1,(0,0),c} \rangle , \ \ \
V_x \mid  \Psi_{nnn,1,(0,0),c} \rangle =- \mid  \Psi_{nnn,1,(0,0),b}
 \rangle , $$
$$ I_- \mid \Psi_{nnn,1,(0,0),d} \rangle =+ \mid  \Psi_{nnn,1,(0,0),d} \rangle , \ \ \
V_x \mid  \Psi_{nnn,1,(0,0),d} \rangle =- \mid  \Psi_{nnn,1,(0,0),a}
\rangle .  $$

\section{Trial states in the $p_y = 0$ case}
We limit ourselves to $S =1$ and $p_x \ne 0$ here. The trial states are
as follows.
\subsection{one-triplet states}
\begin{eqnarray*}
\mid \Psi_{one,1,(p,0),a} \rangle  & \equiv & \mid(p,0)A \rangle
 + \exp(\frac{ip}{2}) \mid(p,0)B \rangle , \\
\mid \Psi_{one,1,(p,0),b} \rangle & \equiv & \mid(p,0)A \rangle
 - \exp(\frac{ip}{2}) \mid(p,0)B \rangle  .
\end{eqnarray*} 
Note that
\begin{eqnarray*}
 V_y \{ \mid(p,0)A \rangle \pm \exp(\frac{ip}{2}) \mid(p,0)B \rangle \}
=\pm \exp(-\frac{ip}{2})
\{ \mid(p,0)A \rangle \pm  \exp(\frac{ip}{2}) \mid(p,0)B \rangle \} \ .
\end{eqnarray*}

\subsection{two-nearest-neighboring-triplet states}
\begin{eqnarray*}
\mid \Psi_{nn,1,(p,0),a} \rangle & \equiv & \mid (p,0) AR \rangle 
+ \exp(\frac{ip}{2})\mid (p,0)BR \rangle ,\\
\mid \Psi_{nn,1,(p,0),b} \rangle & \equiv & \mid (p,0) AR \rangle
 - \exp(\frac{ip}{2})\mid (p,0)BR \rangle ,  \\
\mid \Psi_{nn,1,(p,0),c} \rangle & \equiv & \mid (p,0) AL \rangle
 + \exp(\frac{ip}{2})\mid (p,0)BL \rangle , \\
\mid \Psi_{nn,1,(p,0),d} \rangle & \equiv & \mid (p,0) AL \rangle
 - \exp(\frac{ip}{2})\mid (p,0)BL \rangle  .
\end{eqnarray*}
Note that  
\begin{eqnarray*}V_y \{\mid (p,0) AZ \rangle\pm \exp(\frac{ip}{2})\mid (p,0)BZ \rangle \}
= \pm \exp(-\frac{ip}{2})
\{\mid (p,0) AZ \rangle \pm \exp(\frac{ip}{2})\mid (p,0) BZ \rangle \} ,
\end{eqnarray*}
with $Z= L$ or $Z= R$. 

Also note that $E_{nn,1,(p,0),a} =E_{nn,1,(p,0),c}$ and 
$E_{nn,1,(p,0),b} = E_{nn,1,(p,0),d}$.

\begin{table}[htbp]
   \begin{center}
      \begin{tabular}{lccccc} \hline
trial state $\mid \Psi_m \rangle$& S & $I_+$ & $I_-$  & $U$ & $(E_m-E_0)/J$   \\ \hline  
$\mid \Psi_{nn,0,(0,0),a'} \rangle$ & 0 & $-$ &  $-$ &  $-$ & 0.49 \\
$\mid \Psi_{nn,0,(0,0),b'} \rangle$ & 0 &  $-$ &  $-$ & + & 0.40 \\
$\mid \Psi_{nn,0,(0,0),c'} \rangle$ & 0 &  $-$ & + & $*$ & 0.93 \\
$\mid \Psi_{nn,0,(0,0),d'} \rangle$ & 0 & + &  $-$ & $*$ & 0.93 \\
$\mid \Psi_{nnn,0,(0,0),a'} \rangle$ & 0 & + & + & + & 0.80 \\
$\mid \Psi_{nnn,0,(0,0),b'} \rangle$ & 0 & + & + & $-$ & 0.91 \\ \hline
$\mid \Psi_{nn,1,(0,0),a'} \rangle$ & 1 &  $-$ & $-$  &  $-$ & 0.89 \\
$\mid \Psi_{nn,1,(0,0),b'} \rangle$ & 1 &  $-$ & $-$  & + & 1.04 \\
$\mid \Psi_{nn,1,(0,0),c'} \rangle$ & 1 &  $-$ & + & $*$ & 0.40 \\
$\mid \Psi_{nn,1,(0,0),d'} \rangle$ & 1 & + & $-$  & $*$ & 0.40 \\ \hline
      \end{tabular}      
\caption{The lowest energy  for the trial states with 
different quantum numbers at the coupling ratio $J'/J = 0.65$. 
Asterisks ($*$) in the table denote that the state is not an
    eigenstate of the operator $U$.}
   \end{center}
\end{table}

\begin{figure}[htbp]
\begin{center}
\scalebox{0.5}{\includegraphics{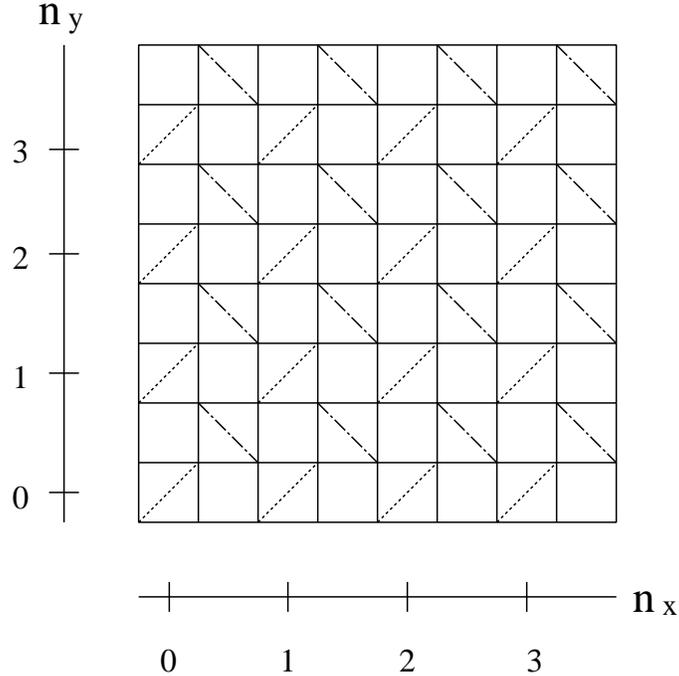}}
\caption{A schematic view of the Shastry-Sutherland model. Each site is
 occupied by a spin. The intra-dimer coupling is denoted by dotted lines
 (A-sublattice) or dot-dashed lines (B-sublattice) and the inter-dimer
 coupling by solid lines.}
\end{center}
\end{figure}

\begin{figure}[htbp]
\begin{center}
\scalebox{0.5}{\includegraphics{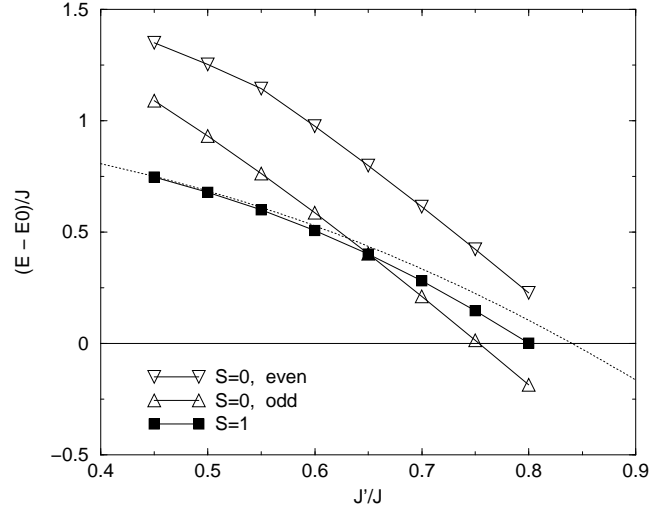}}
\caption{Difference between $E_0= -\frac{3}{8}NJ$
 (the energy of the singlet-dimer
 state) and the lowest energy obtained from the trial states with the 
momentum (0,0). The dotted line is the spin gap
 estimated by the perturbation theory up to the fifth order\cite{fukumoto}, 
$\Delta /J = 1-(J'/J)^2 -\frac{1}{2}(J'/J)^3  -\frac{1}{8}(J'/J)^4 
+\frac{2}{32}(J'/J)^5$.}
\end{center}
\end{figure}

\begin{figure}[htbp]
\begin{center}
\scalebox{0.5}{\includegraphics{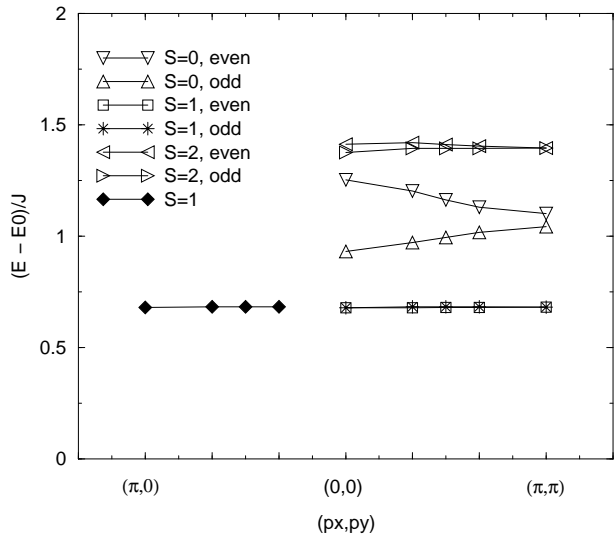}}
\caption{Dispersion of the energy difference 
when the coupling ratio $J'/J = 0.5$.  }
\end{center}
\end{figure}

\begin{figure}[htbp]
\begin{center}
\scalebox{0.5}{\includegraphics{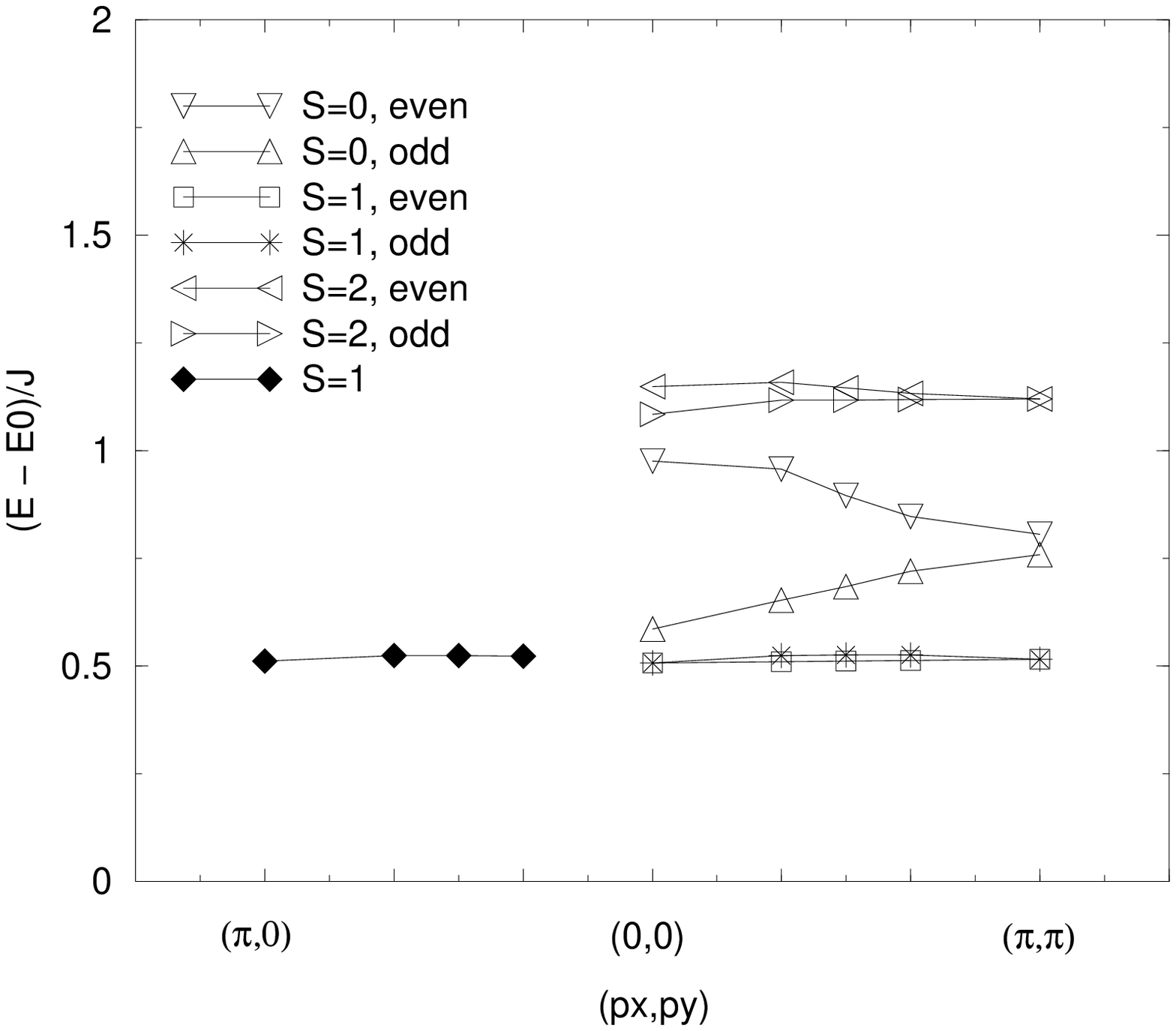}}
\caption{Dispersion of the energy difference when J'/J = 0.6.}
\end{center}
\end{figure}


\begin{figure}[htbp]
\begin{center}
\scalebox{0.5}{\includegraphics{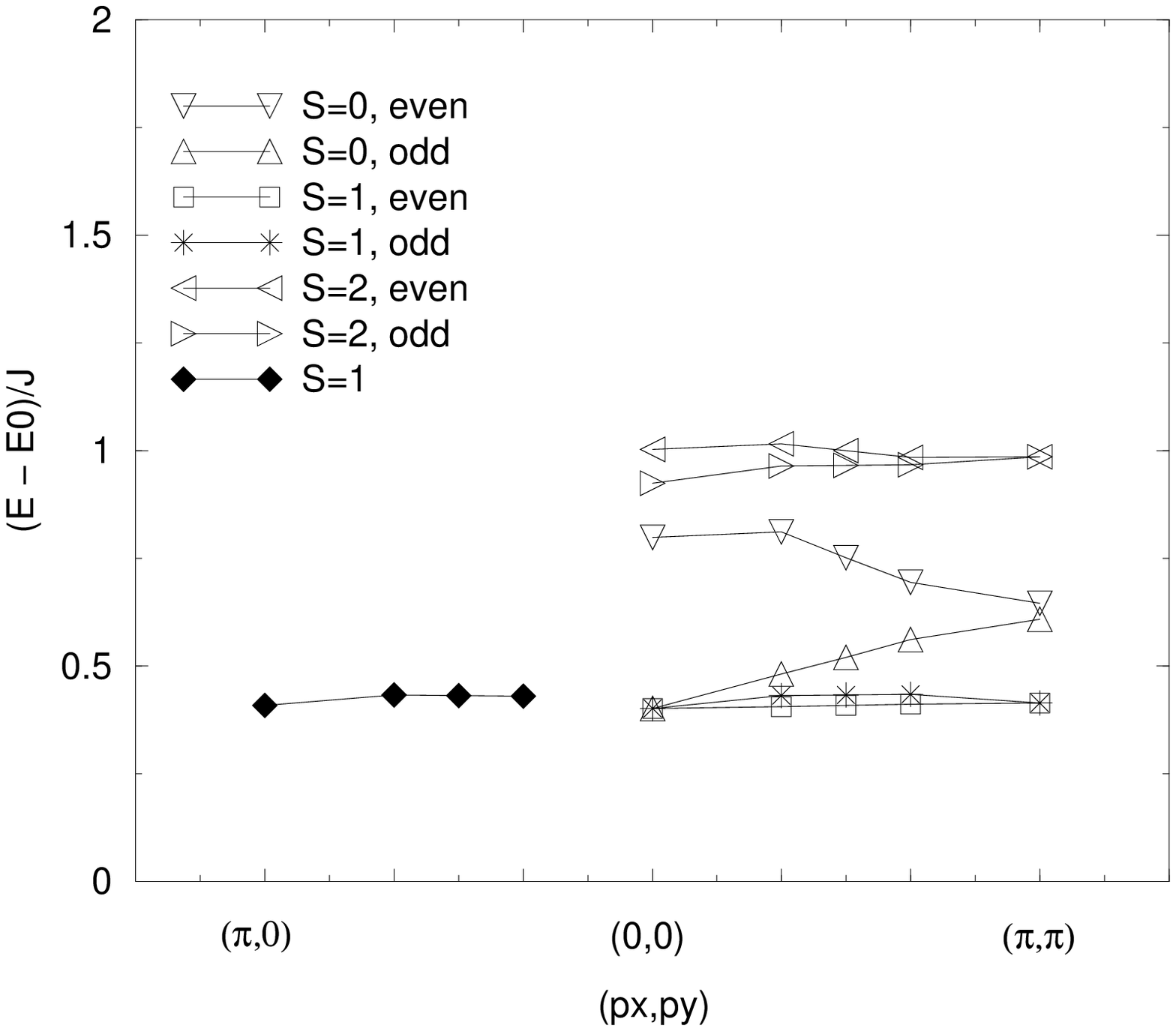}}
\caption{Dispersion of the energy difference when J'/J = 0.65.}
\end{center}
\end{figure}



\begin{thebibliography}{99}
\bibitem{footnote2}
With the order $L_{max}$ higher than 3
one might observe the finite size effects on these lattices.
\bibitem{footnote3}
Note that $r_2$ should be less than
or equal to $r_1$ if there were no helical order phase.
\bibitem{footnote4}
We also
examined the eigenvector of the effective Hamiltonian to make
sure that those components which include the $nnn$-triplets stay dominant.
\end{thebibliography}
\end{document}